\documentclass[letterpaper, 10 pt, journal, twoside]{IEEEtran}  

\IEEEoverridecommandlockouts                              

\usepackage{graphicx} 
\usepackage{amsmath} 
\usepackage{amssymb}  
\usepackage[usenames, dvipsnames]{color}

\usepackage{lipsum}
\usepackage{color}
\usepackage{cite}

\usepackage{diagbox} 
\usepackage{tabu}
\usepackage{balance}  

\usepackage[ruled,linesnumbered]{algorithm2e}
\usepackage{algpseudocode}
\usepackage{epsfig}

\usepackage{booktabs,tabulary,lipsum}
\usepackage{dcolumn,tipa}
\newcolumntype{d}{D{.}{.}{6.5}}
\usepackage{siunitx}
\usepackage[usestackEOL]{stackengine}
\usepackage{multirow}

\newcommand{\revision}[1]{\textcolor{black}{#1}} 
\newcommand{\final}[1]{\textcolor{black}{#1}} 

\title{
Skeleton Graph-based \\Ultrasound-CT Non-rigid Registration 
}

\author{Zhongliang Jiang$^{1}$, Xuesong Li$^{1}$, Chenyu Zhang$^{1}$, Yuan Bi$^{1}$, Walter Stechele$^{2}$, and Nassir Navab$^{1}$, \textit{Fellow, IEEE} 
\thanks{Manuscript received: January, 22, 2023; Revised April, 9, 2023; Accepted May, 9, 2023. This paper was recommended for publication by Editor Jessica Burgner-Kahrs upon evaluation of the Associate Editor and Reviewers' comments. (\textit{Corresponding author:} Zhongliang Jiang.)}
\thanks{$^{1}$Z. Jiang, X. Li, C. Zhang, Y. Bi and N. Navab are with the Chair for Computer Aided Medical Procedures and Augmented Reality, Technical University of Munich, Germany. {\tt\footnotesize{(zl.jiang@tum.de)}}
        }%
\thanks{$^{2}$ W. Stechele is with the Chair of Integrated Systems, Technical University of Munich, 80333 München, Germany}
\thanks{The initial implementation was performed by C. Zhang in his thesis.}
\thanks{Digital Object Identifier (DOI): see top of this page.}
}

\begin{document}

\maketitle


\begin{abstract}
Autonomous ultrasound (US) scanning has attracted increased attention, and it has been seen as a potential solution to overcome the limitations of conventional US examinations, such as inter-operator variations. However, it is still challenging to autonomously and accurately transfer a planned scan trajectory on a generic atlas to the current setup for different patients, particularly for thorax applications with limited acoustic windows. To address this challenge, we proposed a skeleton graph-based non-rigid registration to adapt patient-specific properties using subcutaneous bone surface features rather than the skin surface. To this end, the self-organization mapping is successively used twice to unify the input point cloud and extract the key points, respectively. Afterward, the minimal spanning tree is employed to generate a tree graph to connect all extracted key points. To appropriately characterize the rib cartilage outline to match the source and target point cloud, the path extracted from the tree graph is optimized by maximally maintaining continuity throughout each rib. To validate the proposed approach, we manually extract the US cartilage point cloud from one volunteer and seven CT cartilage point clouds from different patients. The results demonstrate that the proposed graph-based registration is more effective and robust in adapting to the inter-patient variations than the ICP (distance error mean$\pm$SD: $5.0\pm1.9~mm$ vs $8.6\pm6.7~mm$ on seven CTs).  
\end{abstract}

\markboth{IEEE Robotics and Automation Letters. Preprint Version. Accepted May, 2023}
{Jiang \MakeLowercase{\textit{et al.}}: Skeleton Graph-based Ultrasound-CT Non-rigid Registration}


\bstctlcite{IEEEexample:BSTcontrol}
\section{Introduction}

\begin{figure}[ht!]
\centering
\includegraphics[width=0.48\textwidth]{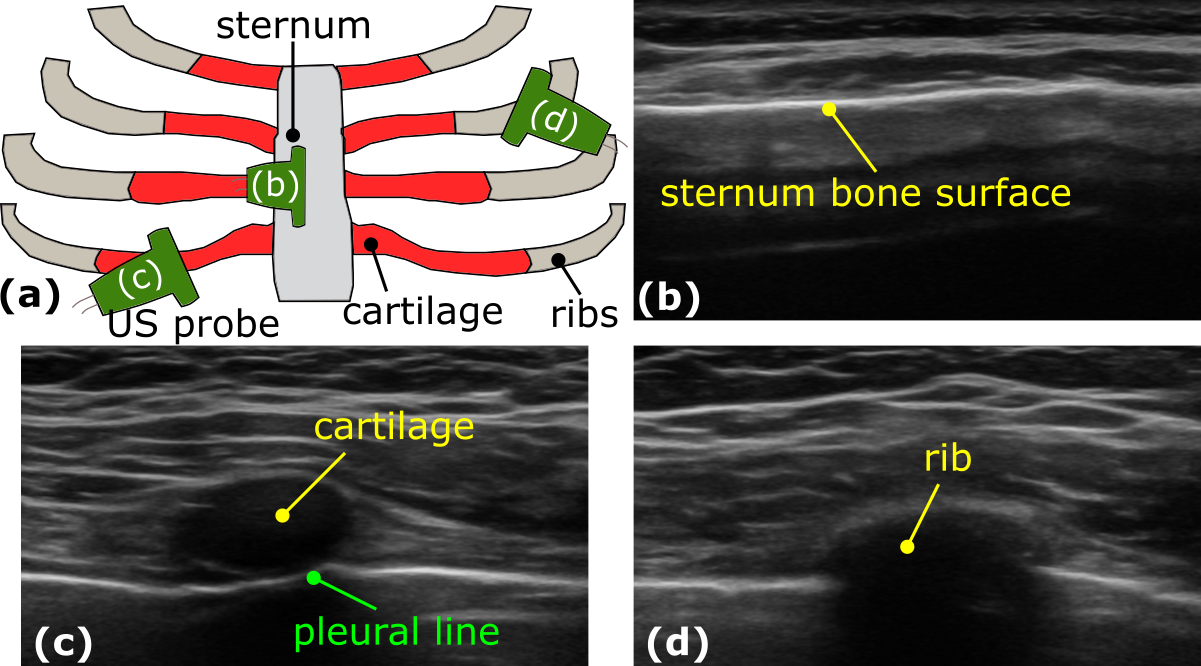}
\caption{Illustration of the anatomical feature of ribs on US images. (a) three types of thorax bones: sternum, rib and costal cartilage. (b), (c) and (d) are the representative B-mode images of sternum, cartilage and rib, respectively. Due to the low acoustic impedance of cartilage, \revision{subcutaneous pleural line can be found right under the cartilage from US images.} 
}
\label{Fig_issue_stament}
\end{figure}

\par
Medical ultrasound (US) is one of the most widely used imaging modalities in daily clinical practice. Since US imaging is real-time, non-ionizing radiation, and widely accessible, it is often used as the standard approach for internal organ examinations~\cite{welleweerd2021out, Jiang2023dopus} and imaging-guided intervention~\cite{napoli2021ultrasound}, particularly in the preliminary healthcare industry. Nevertheless, the conventional US examination is highly user-dependent, as the resulting image is sensitive to the acquisition parameters, i.e., contact force and probe orientation~\cite{gilbertson2015force}. Besides the deep understanding of the physiological background, sonographers are required to simultaneously optimize the acquisition parameters toward high-quality images during US examinations, which often need years of systematic medical training~\cite{maraci2014searching}. 

\par
To address the shortage of experienced sonographers, in particular in rural areas, the robotic technique is employed to develop fully or semi-automatic US screening programs. Due to the superiority in terms of accuracy and repeatability, the robotic US system (RUSS) is considered as a promising solution to overcome the limitation of intra- and inter-operator variations. Pierrot~\emph{et al.} developed a robotic system to assist clinicians in moving a probe on patients' skin~\cite{pierrot1999hippocrate}. They presented in-vivo results for reconstructive surgery to demonstrate the feasibility of applying such a system in real scenarios. Gilbertson~\emph{et al.} employed compliant control strategies to maintain a constant contact force between the probe and the object~\cite{gilbertson2015force}. To accurately control the acquisition parameters, Huang~\emph{et al.} employed a depth camera to compute the normal direction of the target surface~\cite{huang2018robotic}. Besides, Jiang~\emph{et al.} developed a mechanical model-based method to accurately position the probe in the normal direction of unknown constraint surfaces based on contact force~\cite{jiang2020automaticTIE} and real-time US images~\cite{jiang2020automatic}. 
The camera-based approach has high time efficiency, while the force-based approach can achieve high accuracy. Therefore, the former is often used for real-time applications and the latter is more suitable for applications requiring high precision, such as CT-US registration for orthopedics applications~\cite{wein2015automatic}.

\par
To further enable autonomous scans, Huang~\emph{et al.} extract the abdominal area from RGB images based on the rule ``Red$>$Green$>$Blue" and then automatically compute the multiple-line scanning trajectory to fully cover the extracted area~\cite{huang2018robotic}. Jiang~\emph{et al.} calculated the desired scanning direction of the target tubular structure in real-time based on B-mode images~\cite{jiang2021autonomous}. Yet, the proposed optimal control is only validated for tubular structures. Considering the restricted acoustic windows for the organs covered by ribs, e.g., the liver and heart, G{\"o}bl~\emph{et al.} computed scan poses to maintain the coverage of the target on tomographic images and optimize the image quality simultaneously~\cite{gobl2017acoustic}. To transfer the planned trajectory on pre-operative images to the current setup, Hennersperger~\emph{et al.} proposed a skin surface-based registration approach to optimize the transformation between pre-operative images and the current setup captured using an RGB-D camera~\cite{hennersperger2016towards}. Recently, Jiang~\emph{et al.} extended this work by incorporating the non-rigid registration approach to bridge differences between pre-operative MR and individual patients, such as articulated motions in arm scanning~\cite{jiang2022towards}. Both of these approaches are proved to be robust for their specific applications (aorta and limb artery tree). However, since they only use the skin surface extracted from an external RGB-D camera without subcutaneous anatomical features, their use in thorax application is limited, such as the intercostal liver ablation~\cite{quesson2010method}. For such a task, the probe should be accurately positioned between two ribs to guarantee the visibility of the target by avoiding acoustic shadows caused by bone structures (see Fig.~\ref{Fig_issue_stament}).
Without explicitly considering subcutaneous bone structures with high acoustic impedance, the shadows will hinder the visibility of the internal target of interest.

\par
To address this challenge, this work presents a skeleton graph-based non-rigid registration between the cartilage point clouds extracted from a tomographic template and US images of patients. Compared to the existing studies using skin surface point clouds~\cite{hennersperger2016towards, jiang2022towards}, subcutaneous bone can better characterize the scene of thorax examinations, which often require accurate positioning of a probe in limited acoustic windows, i.e., between two ribs. Therefore, the proposed skeleton graph-based registration can more appropriately transfer a planned scanning path to the current setup for robotic scanning. Since there is no camera involved in our setup, the method does not require a hand-eye calibration process, which, as a side benefit, could bring more convenience in clinical practice. The main contributions are summarized as follows:
\begin{itemize}
  \item A skeleton graph-based non-rigid registration is developed to enable the accurate transferring of scanning trajectories from tomographic templates to the current setup based on the invariant subcutaneous bone surfaces. This is particularly useful for thorax applications with limited acoustic windows.
  
  \item Automatically generating skeleton graphs from rib cartilage point clouds by successively using the self-organization map and minimal spanning tree algorithm; then, further optimizing the computed skeletons to match the rib cartilage outline by considering the anatomy's continuity. 
  
  \item The visible cartilage anatomical features on both US and CT images are used as biometric landmarks to enable the registration between a generic CT template and tracked US images characterizing patient-specific properties.
  
  
\end{itemize}
Finally, the proposed graph-based registration is validated on the US data recorded from a volunteer and seven random CT images from a public dataset. The results demonstrate that the proposed method outperforms the classical iterative closest point (ICP) algorithm in terms of adapting to inter-patient variations (distance error mean$\pm$SD: $5.0\pm1.9~mm$ vs $8.6\pm6.7~mm$ on seven CTs). More intuitive results can be found in the online video\footnote{Video: https://www.youtube.com/watch?v=LkSHL7FJ8eU}.

\section{Related Work}
\par
Due to the invariance of bone structures on images, it is often employed as the subcutaneous feature for the registration between pre-operative tomographic images (CT or MR) and inter-operative US images, particularly in orthopedic surgery. The existing approaches can be categorized into two groups: image-based approach~\cite{wein2007simulation} and bone surface-based approach~\cite{wein2015automatic,ciganovic2018registration}. The image-based approaches often directly register US images to the preoperative tomographic images based on the similarity metrics such as LC$^2$~\cite{wein2007simulation}. The 2D-3D registration~\cite{fuerst2014automatic, zheng2017evaluation} has been an attractive direction recently, while it is an ill-posed problem and usually requires rich anatomical features to ensure registration accuracy, like brain images. The image-based approaches have the benefits that there is no need to do bone segmentation, while the result often suffers from the imaging noise~\cite{hacihaliloglu2014local}. 

\par
In contrast, surface-based approaches~\cite{wein2015automatic,ciganovic2018registration} require segmentation results of the subcutaneous anatomy. To segment bone surfaces from US images, various approaches have been proposed in the existing studies~\cite{wein2015automatic, hacihaliloglu2009bone, wang2020robust, alsinan2020bone}. Recently, Li~\emph{et al.} generated synthesized realistic US images from CT scans to facilitate the training of segmentation network~\cite{li2023enabling}. Based on the segmentation results, 3D point clouds can be created by stacking the bone surface with proper tracking information. Based on the point clouds obtained from both modalities, the classical ICP algorithm~\cite{besl1992method} can be used to optimize the transformation matrix between two point clouds. Considering the point clouds may only be partially observed, Zhang~\emph{et al.} incorporate the partially reliable normal vectors to form the registration problem as a maximum likelihood estimation problem~\cite{zhang2021reliable}. The experiments on a femur head showed that the method could robustly and accurately optimize the rigid matrix. But it is also worth noting that the performance may decrease if the target anatomy has limited geometry changes.

\begin{figure*}[ht!]
\centering
\includegraphics[width=0.87\textwidth]{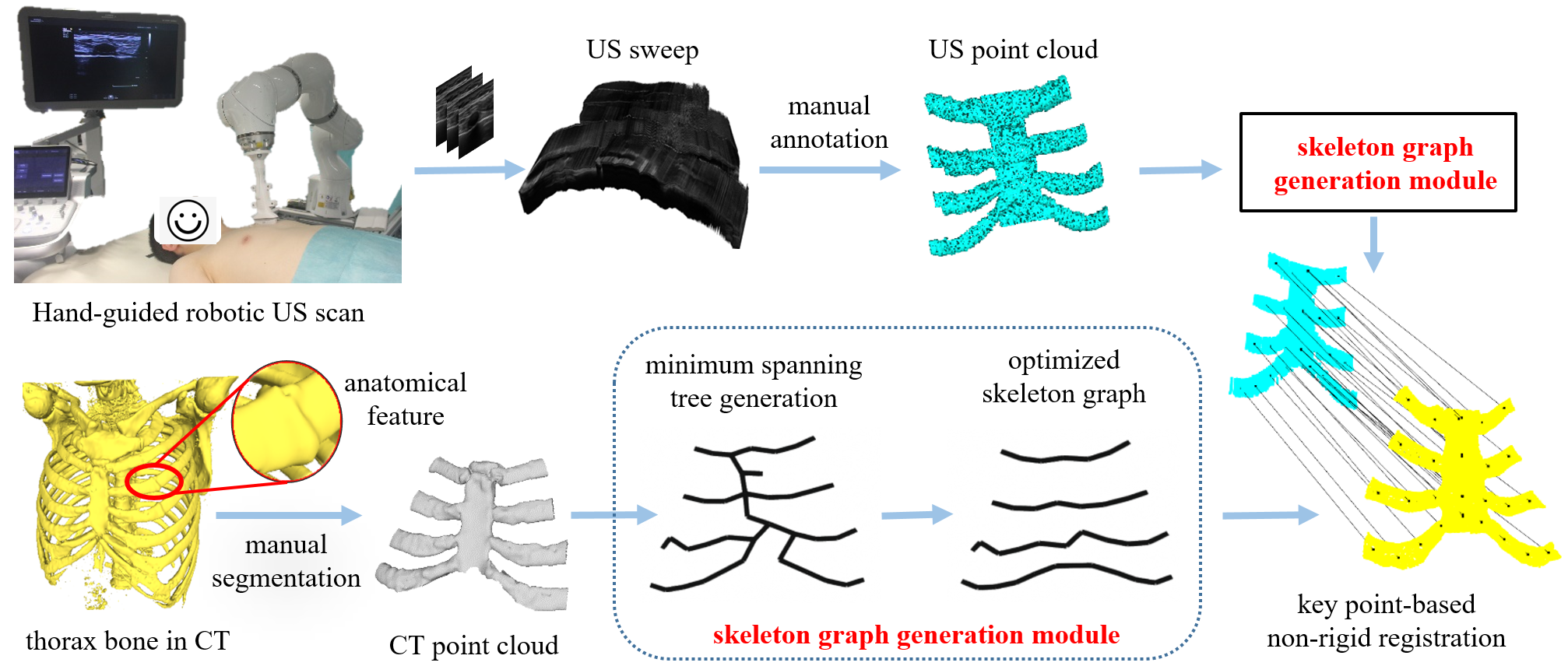}
\caption{The workflow of the proposed skeleton graph-based non-rigid registration framework.}
\label{Fig_workflow}
\end{figure*}

\par
To consider the changes and deformations of targets, Chebrolu~\emph{et al.} segmented the plant leaves and stem, respectively, using a support vector machine and then developed a spatio-temporal non-rigid registration approach~\cite{chebrolu2020spatio, magistri2020segmentation}. In computer vision, the non-rigid ICP~\cite{amberg2007optimal} and the Coherent Point Drift (CPD) algorithm~\cite{myronenko2010point} were proposed for non-rigid point set registration. Such methods often assume that the two inputs are intact and share the same region of interest (ROI), e.g., intact liver or kidney. Specific to the thoracic application with limited acoustic windows, it is hard to guarantee the selected rib ROIs from different patients are consistent with the templates because thoracic ribs cannot be completely scanned only from the front side without moving patients. Therefore, we only count the cartilage part to ensure consistent ROIs for different patients instead of the intact thoracic cage for non-rigid registration in this study. The cartilage's anatomical features on both US and CT images are used as biometric landmarks to enable the registration of the same ROIs from CT templates and individual patients' US scans.

\section{Methods}
The general aim of this work is to develop a skeleton graph-based non-rigid registration approach for autonomous robotic stenography based on the rib cartilage point clouds. The use of subcutaneous bone features enables appropriate transferring of US scanning trajectory from pre-operative tomographic images to the current setup; therefore, allowing autonomous scanning or navigation for thoracic applications with limited acoustic windows, such as liver ablation between ribs~\cite{quesson2010method}. 
The overall workflow of the proposed framework is summarized in Fig.~\ref{Fig_workflow}. We first manually annotate the rib cartilage in both US images and tomographic images based on its visible anatomical characteristic. Then, to appropriately characterize the rib cartilage outline, the self-organizing mapping and minimal spanning tree are successively implemented. Finally, the graph-based non-rigid registration is carried out by matching the key points on the extracted skeleton graph. The details of individual parts are described in the following sections.

\subsection{Robotic US Calibration}
\par
To map US images' pixels into 3D space, spatial US calibration is needed. Besides, it is also important to have an additional temporal calibration between the US stream and robotic tracking data to properly stack the 2D images in 3D space, in particular for the sweep with multiple scan lines. 

\par
\subsubsection{Spatial Calibration}
Spatial calibration refers to the transformations between different coordinate frames. In this study, the involved frames are defined as follows: robotic base $\{b\}$, robotic end-effector $\{ee\}$, tool center point (TCP) $\{tcp\}$ and US image frame $\{us\}$. Similar to~\cite{jiang2021autonomous}, the pixel scale factor is first computed based on the probe footprint length and image depth. Then the transformation from $\{ee\}$ to $\{b\}$ can be obtained based on the given kinematic model from the manufacturer. The final matrix $^{b}_{us}\textbf{T}$ mapping the pixel position from US image $\{us\}$ into the robotic base frame $\{b\}$ can be calculated as follows.

\begin{equation}\label{eq_trasformation}
^{b}_{us}\textbf{T} = ^{b}_{ee}\textbf{T}~^{ee}_{tcp}\textbf{T}~^{tcp}_{us}\textbf{T}
\end{equation}
where $^{j}_{i}\textbf{T}$ is the matrix transferring position representations from frame $\{i\}$ to frame $\{j\}$. 



\subsubsection{Temporal Calibration}
To obtain accurate 3D geometry of the scanned volume, temporal calibration is important for synchronizing the US and robotic tracking streams. Inspired by the temporal synchronization process proposed in~\cite{salehi2017precise}, we repetitively move the robotic upward and downward on a gel phantom with a 3D-printed landmark inside. Then, an intensity threshold-based method is used to extract the landmark location from the recorded B-mode images. Since the image and tracking data are recorded at the same time, the repeated pixel location changes should have the same change tendency as the robotic position, e.g., the motion in the probe centerline direction. Thereby, the temporal difference can be computed by maximally matching the two curves with respect to time stamps. The temporal synchronization shift between the US and robotic tracking streams is $268~ms$ in our setup. 

\begin{figure*}[ht!]
\centering
\includegraphics[width=0.98\textwidth]{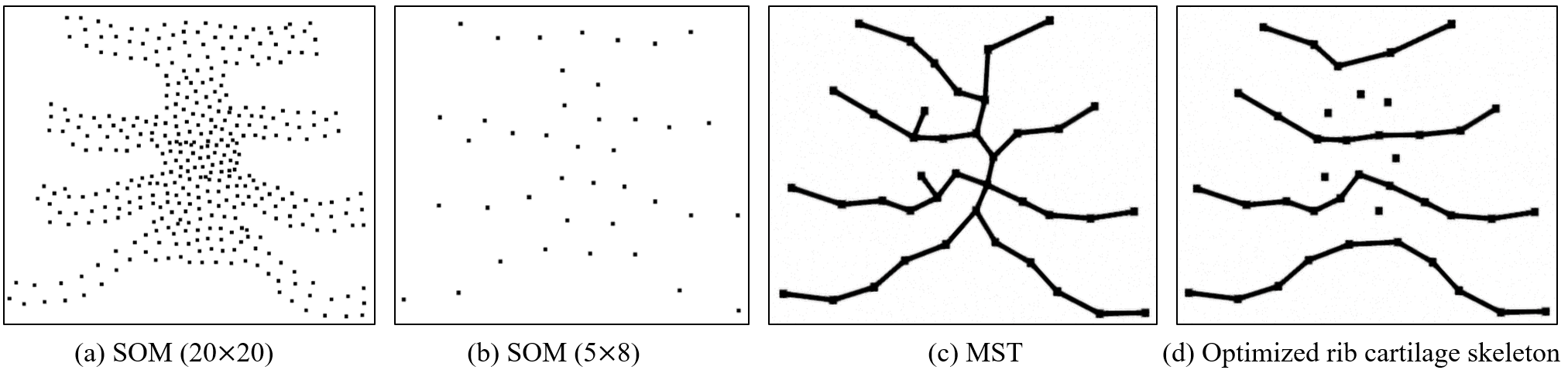}
\caption{Rib cartilage skeleton graph extraction and optimization from the US point cloud $\textbf{P}_{us}$. (a) the key points $\textbf{P}^{som}_{us}$ extracted form original $\textbf{P}_{us}$ using a SOM with a $20\times20$ grid map. (b) the resulting key vertex points $\textbf{V}_p^{us}$ obtained from $\textbf{P}^{som}_{us}$ after applying a SOM with a $5\times8$ grid map. (c) the MST $G_{tr}$ computed based on the key points extracted from $\textbf{V}_p^{us}$. (d) the optimized four paths representing four rib cartilage. 
}
\label{Fig_skeleton}
\end{figure*}

\subsection{Cartilage Point Cloud Generation} \label{sec:III_Point_Cloud_Generation}
\par
The selection of consistent ROI is one of the practical problems that limit the application of non-rigid registration. Due to inter-patient variations, it is hard to accurately select the same ROIs from different patients in different imaging modalities. Inspired by the finding that the cartilage US images are different from ribs on B-mode images~\cite{van2019dynamic}, where we can also observe pleural boundaries under rib cartilage (see Fig.~\ref{Fig_issue_stament}), we can select common cartilage parts from different patients. Besides, there are also apparent anatomical features existing at the connection part between cartilage and ribs (see Fig.~\ref{Fig_workflow}) in CT images. Therefore, we can get complete cartilage both in tomographic images and US images. Based on this, a non-rigid registration developed based on cartilage bones becomes feasible and such skeleton graph-based registration can enable transferring of a planned path from pre-operative images to the current setup of various patients.

\subsubsection{Point Clouds Generation from CT Template}
\par
Seven chest templates are selected from a public human CT dataset\footnote{CT dataset: https://github.com/M3DV/RibSeg}. To extract the ribs from CT data, we first perform an intensity-based segmentation in an open-source software 3D Slicer\footnote{3D Slicer: https://www.slicer.org/}. A representative result of the chest mesh is depicted in Fig.~\ref{Fig_workflow} (bottom left part). Based on the anatomical features, we manually segment the cartilage sections from the $2$-th, $3$-th, $4$-th and $5$-th ribs in Meshlab\footnote{Meshlab: https://www.meshlab.net/}. To create uniformly distributed cartilage point clouds $\textbf{P}_{ct}$ from CT, the Poisson disc sampling algorithm is employed instead of the random sampling method.


\subsubsection{Point Clouds Generation from US Sweep}
\par
To generate the subcutaneous bone surface point clouds, we first ask the volunteers to lie down on a flat operation bed as Fig.~\ref{Fig_workflow} (top left part). Then, a human operator manually defines a multiple-line US scanning to cover the upper part of the chest. The US acquisition took $106~s$ on average when the volunteer was under natural breathing. Based on the US calibration results, the 3D visualization of the recorded sweep can be seen in Fig~\ref{Fig_workflow}. To further extract the cartilages point cloud $\textbf{P}_{us}$, we first label the cartilage bone surface by manually annotating the bone surface on 2D B-mode images using ImFusionSuit (ImFusion AG, Germany). Afterward, the original B-mode images are replaced with the manually annotated binary masks of cartilage bone to form the 3D object and further compounded using a linear interpolation using PLUS~\cite{lasso2014plus}. Then, the same processes described in the previous subsection are performed in Meshlab to generate US cartilage point cloud (including sternum) $\textbf{P}_{us}$ (see Fig.~\ref{Fig_workflow}).


\subsection{Skeleton Graph Generation and Optimization}

\subsubsection{Self-Organizing Map for Data Cleaning and Key Point Extraction}
\par
The self-organizing map (SOM) algorithm~\cite{kohonen1990self} is proposed by Kohonen to represent high-dimensional features using a lower dimensional representation while maintaining the topological structure. It was used to extract plant's key points in~\cite{magistri2020segmentation}. Inspired by this work, we apply SOM to extract the key points from CT and/or US rib cartilage point clouds. 
The SOM is an unsupervised machine learning technique trained using competitive learning, which ensures it can be used on unseen data. To extract the point cloud distributed uniformly in Euclidean space using the SOM algorithm, a grid map $G_{g}$ with size of $m_g\times n_g$ is initialized. Each node of $G_{g}$ contained a corresponding weight vector $\textbf{W}_{s}$ with the same dimension as input data $\textbf{P}$.
To update the weight vectors $\textbf{W}_{s}$, the Euclidean distance between a random sample of the training input and all nodes of $G_{g}$ are computed. The weight vector with the smallest value is called the best matching unit (BMU). The node close to the BMU will allow larger adjustments than the ones far away from BMU. Afterward, $\textbf{W}_{s}$ is updated as follows:

\begin{equation}\label{eq_som}
W_{s}(i+1)=W_{s}(i)+\theta_{(BMU,i)}\cdot l_r\cdot \left[\textbf{P}(k)-W_{s}(i)\right]
\end{equation}
where $i$ is the current iteration, $\theta_{(BMU,i)}$ is the updated restriction functions computed based on the grid distance between other nodes and the BMU, $l_r$ is the learning rate, and $\textbf{P}(k)$ is the $k$-th data in the input source point cloud. 

\par
In this study, the SOM is successively applied twice. The first SOM is used to get a uniformly distributed point cloud with a smaller size than the original input point cloud. Furthermore, this process can create point clouds (from CT, US, or different patients) including a unified number of points. Therefore, a unified input has the potential to benefit the generalization capability of the proposed graph-based registration against inter-patient variation. To preserve the character of the original skeleton point cloud, a relatively large number of map nodes ($400=20\times 20$) are used to unify the size of input point clouds. The second SOM is used to automatically extract the key skeleton points to represent the rib cartilage point cloud. Since the point clouds are quasi-symmetrical about the sternum, the second map size is set around $m_g\times 2m_g$. Since the point cloud includes four cartilage, $m_g$ is set to $4$. By testing a few paired $m_g$ and $n_g$ around ($4\times8$), we empirically use $5\times8$ for the initial map of the second SOM based on the extracted cartilage skeleton profiles. The representative results of the two SOM results on a US rib cartilage point cloud are depicted in Fig.~\ref{Fig_skeleton} (a) and (b), respectively.

\subsubsection{Minimum Spanning Tree Generation}
\par
After obtaining a set of skeleton key points $\textbf{V}_p$ [Fig.~\ref{Fig_skeleton} (b)], the minimum spanning tree (MST) is used to connect the skeleton key points to form a representative tree characterizing the feature of rib cages. The MST is a graphical generation approach to connect $\textbf{V}_p$ by edges. The spanning tree means a connected graph including all $\textbf{V}_p$, but does not have any closed loops. Here, Euclidean distance is employed as the edge weights for two arbitrary points. The weight matrix $\textbf{W}$ is defined as follows:

\begin{equation} \label{eq_weight_matrix}
\textbf{W}_{i,j}=\Vert \textbf{V}_p(i) - \textbf{V}_p(j) \Vert
\end{equation}
where $i$ and $j$ are the $i$-th and $j$-th point in $\textbf{V}_p$, respectively.  

\par
In order to find the tree graph with a minimum total weight for an undirected graph, Prim's algorithm~\cite{prim1957shortest} is used. This greedy algorithm builds MST $G_{tr}$ step by step from an arbitrary point. In each step, $G_{tr}$ is augmented by adding the cheapest possible edge connection from the tentative tree $G_{tr}$ to the reaming points not yet in $G_{tr}$. A representative MST for the collected cartilage point cloud $\textbf{P}_{us}$ is depicted in Fig.~\ref{Fig_skeleton}~(c).

\subsubsection{Skeleton Tree Optimization}
\par
It can be seen from Fig.~\ref{Fig_skeleton}~(c) that the generated MST $G_{tr}$ connected all $\textbf{V}_p$. Besides, the path connecting two arbitrary nodes in $G_{tr}$ is unique. Since there are four ribs ($2$-th, $3$-th, $4$-th and $5$-th) involved in our setup, we can obtain four representative and unique paths to represent the four ribs. To extract such paths, two endpoints of each rib need to be identified. To this end, the gift wrapping algorithm~\cite{preparata1977convex} is employed to extract the convex hull for the $G_{tr}$ of cartilage point cloud $\textbf{P}_{us}$ and $\textbf{P}_{ct}$. The eight endpoints $\textbf{P}_{end}\in R^{8\times1\times3}$ of the four ribs are the key points forming the convex hull. To further understand the correspondence between the identified eight points in US data, we match $\textbf{P}_{end}^{us}$ to the $\textbf{P}_{end}^{ct}$ using the ICP algorithm. 
According to the ribs order information on CT template, $\textbf{P}_{end}$ can be grouped into the pairwise endpoints set $\textbf{P}_{pair}\in R^{4\times2\times3}$. 

\par
Based on the ordered four pairs of endpoints $\textbf{P}_{pair}$, four unique paths can be obtained from the MST tree by finding the shortest tree distance. Due to the fact that the human ribs are continuous, the direction variations between the vectors connecting two neighboring key points should be small. Therefore, four rib skeleton representations $\textbf{P}_{rs}$ are optimized as follows:

\begin{equation} \label{eq_skeleton_optimiztion}
\textbf{T}_{rs}^{i}=\begin{cases}
               \left[\textbf{T}_{rs}^{i}, \textbf{P}_{rs}^{i}(j)\right] & \text{if}~\angle\theta \leq T_{\theta}  \\
               \textbf{T}_{rs}^{i} & \text{otherwise}\\
            \end{cases}
\end{equation}
where $i=\{2,3,4,5\}$ represents the $i$-th rib cartilage skeleton, $\textbf{T}_{rs}^{i}$ is the optimized rib skeleton, which is initialized as $\textbf{T}_{rs}^{i}(1) = \textbf{P}_{rs}^{i}(1)$, $j = 3,...,N_{i}$ is the $j$-th point of $i$-th rib skeleton, \revision{$\angle\theta = \angle \left(\overrightarrow{\textbf{T}_{rs}^{i}(j-2)\textbf{T}_{rs}^{i}(j-1)},~ \overrightarrow{\textbf{T}_{rs}^{i}(j-1)\textbf{P}_{rs}^{i}(j)}\right)$}, $T_{\theta}$ is the angular threshold filter out unsuitable points. \revision{$T_{\theta}$ is empirically set to $60^{\circ}$ based on the experimental results.} The optimized $\textbf{T}_{rs}$ are visualized in Fig.~\ref{Fig_skeleton}~(d).

\subsection{Skeleton Graph-based Non-rigid Registration}
\par
Based on the extracted skeleton in Fig.~\ref{Fig_skeleton} (d), the four rib cartilage are characterized. To perform the graph-based registration between the US and CT cartilage point clouds ($\textbf{P}_{us}$ and $\textbf{P}_{ct}$), the skeleton key points are fitted using a cubic polynomial function and then resampled uniformly for each rib. Considering the variation in the length, different sampling points are used for different ribs, while the same number of key points will be used for the same rib (e.g., the second rib) in CT and US. The resampled points on the skeleton are computed as follows: 

\begin{equation} \label{eq_resample}
[X^{i}, Y^{i}, Z^{i}] = \left[X^{i}, f_y^i(X^{i}), f_z^i(X^{i})\right]
\end{equation}
where $f_y^i$ and $f_z^i$ are the fitted cubic polynomial functions with respect to X in Y and Z directions, respectively. $i=\{2, 3, 4, 5\}$ is the level of ribs, $X^{i}\in R^{N_i\times1}$ is the $N_i$ values uniformly sampled in X direction with equal interval. Based on the length of each rib cartilage, $N_2$, $N_3$, $N_4$, and $N_5$ are empirically set $6$, $6$, $8$ and $10$, respectively. 

\par
After obtaining the re-sampled skeleton points, the graph-based non-rigid registration can be implemented by computing local transformations for individual points in point clouds. To this end, the closest $N_r$ re-sampled skeleton key points to the specific point in $\textbf{P}_{ct}$ are automatically selected based on Euclidean distance. Then, a local transformation matrix $^{us}_{ct}\textbf{T}$ mapping $\textbf{P}_{ct}$ to $\textbf{P}_{us}$ can be optimized by minimizing Eq.~(\ref{eq_rotatopn_optimization}).

\begin{equation} \label{eq_rotatopn_optimization}
\min_{^{ct}_{us}\textbf{T}} \frac{1}{N_r}\sum_{i=1}^{N_r}{||^{us}_{ct}\textbf{T}~^{ct}P_i - {^{us}P_i}||^2}
\end{equation}
where $^{us}P_i$ and $^{ct}P_i$ are the paired re-sampled skeleton key points of US and CT point clouds, respectively. 
The hyperparameter $N_r=10$ is empirically determined based on the experimental results by making a compromise between anatomy continuity and local accuracy. A large $N_r$ will reduce the non-rigid probability. An extreme case is that if we use all key points to calculate the transformation matrix, the method will be degraded to a rigid registration.


\section{Results}

\subsection{Robotic US System}
\par
The presented RUSS consists of a robotic manipulator (LBR iiwa 7 R800, KUKA GmbH, Germany) and an ACUSON Juniper US System (Siemens Healthineers, Germany). A linear probe (12L3, Siemens, Germany) is attached to the robot flange by a 3D printed holder (see Fig.~\ref{Fig_workflow}). The robot status is updated at $100~Hz$. The US images are captured by a frame grabber (Epiphan video, Canada) in $30~fps$. To properly visualize the bone structure in B-mode images, a default setting provided by the manufacturer is used in this study: MI: $1.13$, TIS: $0.2$  TIB: $0.2$ DB: 60 dB. Since the ribs of interest are shallow, the imaging depth was set to $35~mm$. 

\subsection{Performance of the Skeleton Graph-based Registration}
\par
To quantitatively evaluate the proposed skeleton graph-based non-rigid registration, seven human chest CT images are selected from a public dataset. Using multiple images is helpful in drawing the correct conclusion by ruling out occasional cases. Following the steps described in Sec.~\ref{sec:III_Point_Cloud_Generation}, seven CT point clouds can be obtained to evaluate the registration performance in different cases. The US point cloud is obtained from a healthy volunteer (gender: male, year: $26$, BMI: $19.8$, height: $174~cm$).

\par
\begin{figure}[ht!]
\centering
\includegraphics[width=0.42\textwidth]{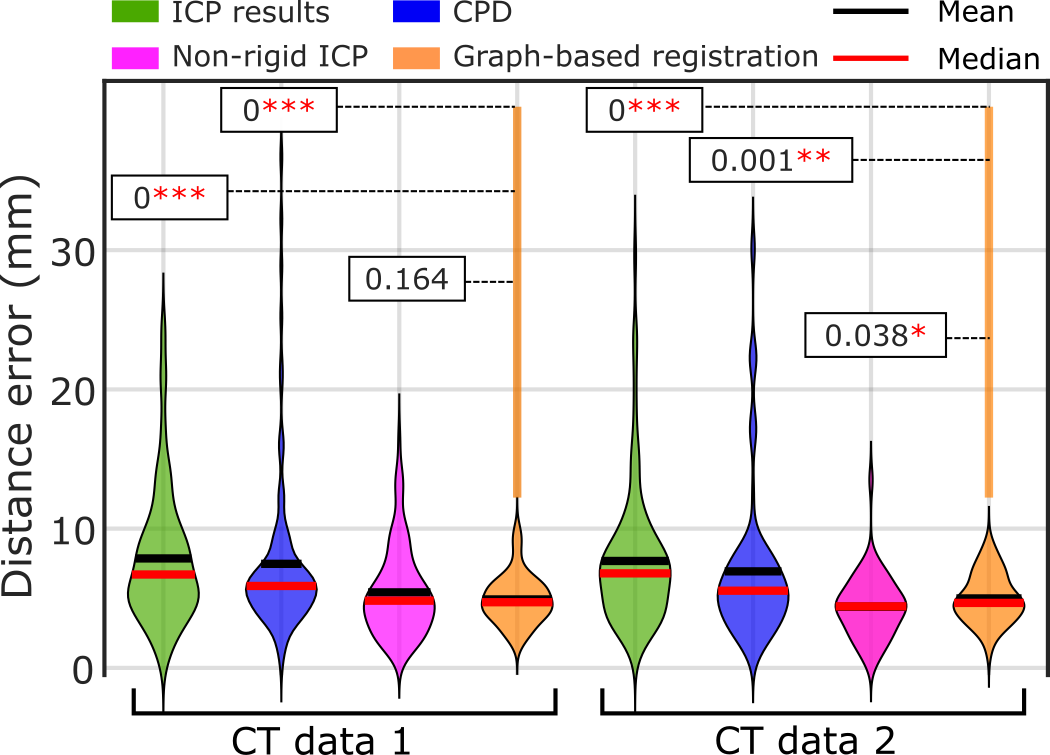}
\caption{The quantitative results of the existing standard ICP, CPD and non-rigid ICP algorithm and the proposed skeleton graph-based non-rigid registration method. The Euclidean distance is used as the validation metric. The p-values are computed between the results computed using the proposed and existing methods. \final{* ($p<0.05$), ** ($p<0.01$) and *** ($p<0.001$) mean there are significant differences.
}}
\label{Fig_result_violin}
\end{figure}

To further compare the performance of the proposed approach to existing ones, i.e., standard ICP, non-rigid ICP, and CPD algorithm, the computed registration errors on \final{two representative CTs} are depicted in Fig.~\ref{Fig_result_violin}. Since the ICP and non-rigid ICP are sensitive to the initial alignment, we manually adjust the source and target point clouds to be close and also align them in the right orientation in order to provide a fair comparison. The proposed skeleton graph-based approach does not require dedicated initialization. The final results are computed in terms of Euclidean distance, which is calculated between a point in the realigned $\textbf{P}^{'}_{ct}$ (after registration) and the closest point in $\textbf{P}_{us}$. It can be seen from Fig.~\ref{Fig_result_violin} that the proposed method can significantly outperform standard ICP and CPD in terms of distance error. The mean ($\pm SD$) error of standard ICP, CPD, and the proposed method are $7.9\pm4.8~mm$, $7.5\pm6.0~mm$, and $4.9\pm1.8~mm$ on CT 1 and $7.7\pm4.9~mm$, $6.9\pm6.0~mm$, and $5.0\pm 1.4~mm$ on CT 2. 
\final{To perform the t-test for statistical comparison, the Lilliefors test is employed to confirm that the estimated errors are normally distributed.}
The $p$-value for the results computed using the proposed method and the standard ICP and CPD are $3.4\times10^{-8}$, and $6.0\times10^{-5}$ on CT 1, and $5.1\times10^{-7}$ and $1.1\times10^{-3}$ on CT 2, respectively. The rigid ICP algorithm performs worst because the inter-patient variations are not taken into consideration. Based on the data distributions, the results computed using CPD are more concentrated than the standard ICP. \final{Additionally, the non-rigid ICP and the proposed method can achieve similar performance (CT 1: $5.4\pm3.2~mm$ vs $4.9\pm1.8~mm$, and CT 2: $4.4\pm2.2~mm$ vs $5.0\pm 1.4~mm$). However, it is worth noting that the proposed method can limit the maximum error ($10.2~mm$ vs $16.4~mm$ on CT 1 and $9.3~mm$ vs $13.5~mm$ on CT 2), and its performance is invariant to the initial alignments.}


The Euclidean distance (ED) between two point sets can not prevent the self-deformation for non-rigid ICP. To ensure the original shape can be preserved after non-rigid registration, we compute the Hausdorff distance (HD) between the transferred CT and the US point clouds. The HD is often used to describe how far two unpaired subsets are, which is sensitive to the boundary shape. \final{The statistical results in terms of ED and HD using seven independent CT data are summarized in TABLE~\ref{tab1_comparsion}.} It can be seen that both ED and HD errors achieved by the proposed method are smaller and more stable than others over different CTs. The small ED error and large HD results achieved by non-rigid ICP on CT 1, 2, and 4 indicate self-deformation. After applying the variable transformation matrix to individual CT points, multiple CT points are mapped to the same US points. The maximum HD computed by the proposed method is $13.7~mm$, whereas the one computed by the non-rigid ICP is $69.4~mm$. Therefore, we consider the proposed graph-based non-rigid registration is more effective and robust in adapting to inter-patient variations.

 \begin{table*}[ht]
 \sisetup{
 table-number-alignment = center,
 table-figures-integer = 1,
 table-figures-decimal = 4
 }
 \begin{center}
 \caption{Registration Results of Different Approaches (mean)
 }
 \centering
 \renewcommand\footnoterule{\kern -1ex}
 \renewcommand{\arraystretch}{1.3}
 \resizebox{0.87\textwidth}{!}{
 \revision{
 \begin{tabular}{l*{14}{S}}
 \toprule \multirow{2}{*}{Methods}   & 
 \multicolumn{2}{c}{CT 1} &   
 \multicolumn{2}{c}{CT 2} &
 \multicolumn{2}{c}{CT 3} &
 \multicolumn{2}{c}{CT 4} &
 \multicolumn{2}{c}{CT 5} &
 \multicolumn{2}{c}{CT 6} &
 \multicolumn{2}{c}{CT 7}  \\   
 \cmidrule(r){2-15}  &
 \multicolumn{1}{c}{ED} & 
 \multicolumn{1}{c}{HD} & 
 \multicolumn{1}{c}{ED} & 
 \multicolumn{1}{c}{HD} &
 \multicolumn{1}{c}{ED} & 
 \multicolumn{1}{c}{HD} & 
 \multicolumn{1}{c}{ED} & 
 \multicolumn{1}{c}{HD} &
 \multicolumn{1}{c}{ED} & 
 \multicolumn{1}{c}{HD} & 
 \multicolumn{1}{c}{ED} & 
 \multicolumn{1}{c}{HD} & 
 \multicolumn{1}{c}{ED } & 
 \multicolumn{1}{c}{HD}  \\\midrule 
 ICP\cite{besl1992method} $\clubsuit$ & {7.9} & {59.3} & {7.7} & {32.5} & {7.1} & {31.2} & {6.2} & {15.1} & {16.3} & {68.8} & {9.8} & {43.0} & {5.3}  & {21.6} \\
 CPD\cite{myronenko2010point} & {7.5} & {36.7} & {6.8} & {30.2} & {6.7} & {35.8} & {8.3} & {38.5} & {6.0}  & {25.2} & {7.9} & {46.1} & {9.5}  & {56.7} \\
 Nonrigid ICP\cite{amberg2007optimal} $\clubsuit$  & {5.4} & {69.4} & {\textbf{4.4}} & {34.1} & {\textbf{3.9}} & {\textbf{10.3}} & {\textbf{4.1}} & {29.9} & {\textbf{3.7}}  & {15.8} & {\textbf{3.7}} & {19.7} & {\textbf{4.1}}  & {16.2} \\
 Our method     & {\textbf{4.9}} & {\textbf{11.7}} & {5.0} & {\textbf{12.8}} & {4.9} & {10.7} & {5.7} & {\textbf{13.7}} & {4.9}  & {\textbf{13.2}} & {4.9} & {\textbf{10.5}} & {4.7} & {\textbf{9.8}} \\\bottomrule
 \multicolumn{15}{l}{ED: Euclidean distance; HD: Hausdorff distance; $\clubsuit$: sensitive to initial alignments;  Unit: mm}
 \end{tabular}
 }
 }
 \label{tab1_comparsion}
 \end{center}
 \end{table*}

\subsection{Validation of Transferring Trajectory from CT to US}
\par
\begin{figure}[ht!]
\centering
\includegraphics[width=0.47\textwidth]{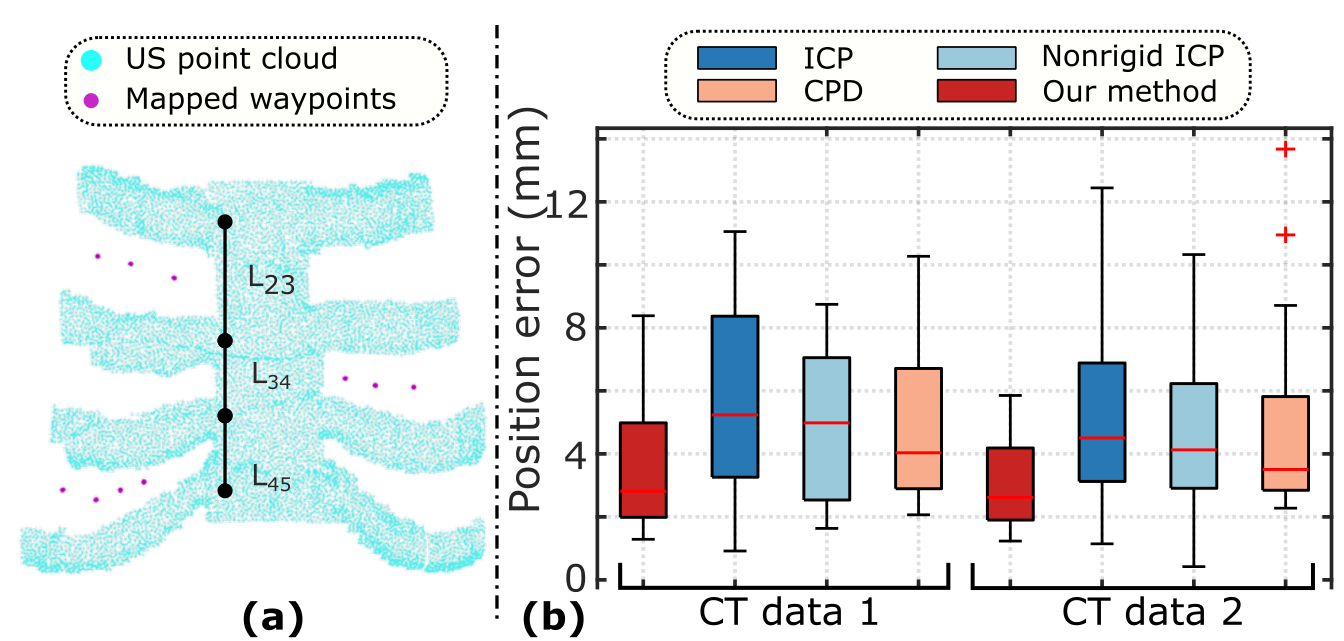}
\caption{Results of trajectory transferring from CT to US spaces. (a) Illustration of the results of $10$ waypoints mapped from CT to US. (b) The statistical results of the mapping performance in terms of position errors obtained by various methods. The length of the neighbouring ribs are measured as $L_{23} = 31.6~mm$, $L_{34} = 26.1~mm$ and $L_{45}=24.2~mm$. }
\label{Fig_path_transfer}
\end{figure}

In this section, we further validate whether the proposed graph-based registration can appropriately map a planned trajectory from generic template CT to the current setup. To this end, we manually determined ten scanning waypoints in both $\textbf{P}_{us}$ and $\textbf{P}_{ct}$. The positions of the ten waypoints are defined by computing the average of the re-sampled key points [Eq.~(\ref{eq_resample})] in two neighboring ribs. It can be seen from Fig.~\ref{Fig_path_transfer} (a), that the ten waypoints are located in the middle of two ribs. To avoid biased results, the ten points are defined in all three levels of the intercostal gaps, which are distributed on both sides. These ten points were only used for evaluating the trajectory transferring performance.

\par
A representative result is intuitively described in Fig.~\ref{Fig_path_transfer} (a), where the red points are the transferred points from the CT to the US point cloud obtained from a volunteer. The transferred points are roughly distributed in the middle of ribs in US point cloud as expected. To further provide quantitative analysis on trajectory transferring performance, the statistical errors are summarized in Fig.~\ref{Fig_path_transfer} (b). The position error is represented by the distance difference between the transferred points from CT and the points defined on US point clouds directly.

\par
The trajectory transferring results obtained by different methods are consistent between the two representative CT images. Compared to existing methods, the errors obtained by the proposed method are the smallest in both cases (mean$\pm$ SD) $3.4\pm1.9~mm$ and $3.1\pm1.5~mm$ on CT 1 and 2. The results obtained by the standard ICP, non-rigid ICP and CPD are $5.6\pm3.0~mm$, $5.0\pm2.5~mm$, and $4.9\pm2.5~mm$ on CT 1 and  $5.2\pm3.0~mm$, $4.9\pm2.7~mm$, and $5.0\pm3.1~mm$ on CT 2. It is also noteworthy that the standard ICP performs worst in terms of positional error in both cases. The two non-rigid methods achieve similar accuracy. Considering the gap between two ribs centerlines varies from $24.2~mm$ to $31.6~mm$ in our setup, which is more than six times larger than the positional error. 
Since the average intercostal space is around $15~mm$~\cite{kim2014sonographic} and the footprint's width of the used probe is around $10~mm$, the tolerance of positioning error will be around $5~mm$. 
Therefore, we consider the proposed graph-based approach can meet the clinical requirement and can be used to remap the planned scan trajectory to current setups for autonomous robotic scanning.

\par
\section{Discussion}
The proposed skeleton graph-based non-rigid registration has been validated on the cartilage point clouds obtained from a volunteer and seven CT images. The results demonstrate that the proposed approach can achieve better performance than the standard ICP, CPD~\cite{amberg2007optimal} and nonrigid ICP~\cite{myronenko2010point} algorithms in adapting to inter-patient variations. However, there are still some limitations that need to be discussed to further inspire future research. First, the physiological motion, i.e., respiration, has not been explicitly considered, which will affect the accurate representation of the bone surface in 3D. To alleviate this issue, future work can try to estimate the respiration-induced motion using external tracking~\cite{jiang2020model} or real-time B-model images~\cite{dai2021deep}.
Second, the skin surface hasn't been included in the registration approach, which can result in undesired contact conditions (too deep or with a gap) between the probe and the patient's skin. However, this negative influence could be reduced by using a compliant control~\cite{jiang2020automaticTIE, gilbertson2015force, welleweerd2021out} to maintain a contact force during scans. Besides, due to the physiological variations between males and females, the breast will result in significant deformation during the acquisition of US images for females. Such inevitable deformation will bring challenges in both acquiring accurate point clouds of the subcutaneous bone surface and appropriately pointing a probe to visualize the internal objects. 

\par
After validating the feasibility of explicitly using cartilage bone for trajectory mapping, future studies need to reduce the manual component in the current pipeline to improve the level of autonomy. To autonomously plan a scanning path for covering the chest part, an external depth camera can be used to extract the object surface based on predefined rules~\cite{huang2018robotic, jiang2022towards} or learning-based segmentation methods~\cite{jiang2022precise}. To ensure the coverage of the whole ROI, Tan~\emph{et al.} employed multiple cameras and fused the surface point clouds obtained from different views~\cite{tan2022fully}. In addition, the US cartilage surface point was obtained by manual annotation based on the anatomical feature depicted in Fig.~\ref{Fig_issue_stament}. Future studies need to further integrate advanced bone surface extraction methods~\cite{wein2015automatic, hacihaliloglu2009bone, wang2020robust, alsinan2020bone, salehi2017precise} to improve the autonomy level. Furthermore, considering practical issues in real scenarios, potential tissue motions~\cite{jiang2021motion, jiang2022precise} and deformations~\cite{jiang2021deformation} should be properly compensated for to ensure acquisition robustness and anatomy accuracy.

\section{Conclusion}
This work presents a skeleton graph-based non-rigid registration approach between tomographic templates and B-mode images of patients for autonomous transferring of US scan trajectory from CT to the current setup. Compared to the existing studies using skin surface point clouds~\cite{hennersperger2016towards, jiang2022towards}, subcutaneous bone surfaces are used to better characterize challenging scenes of thorax application with limited acoustic windows; therefore achieving bone-aware registration results. The proposed method has been validated on the US data recorded from a volunteer and seven CT volumes. The registration results shown in Fig.~\ref{Fig_result_violin} and TABLE~\ref{tab1_comparsion} demonstrate that the proposed method outperforms the standard ICP and CPD algorithms in terms of adapting to inter-patient variations. Although the non-rigid ICP method obtains similar ED results, the proposed method can achieve significantly lower HD results than non-rigid ICP. This indicates that the skeleton graph-based method can better preserve the topology structure of the anatomies, which can improve the robustness and time efficiency of the graph-based registration by leveraging prior skeleton knowledge. By explicitly considering subcutaneous bone structures, we believe this study can aid in the development of autonomous RUSS scanning or navigation for challenging thoracic applications with limited acoustic windows, e.g., liver ablation through intercostal space.



\bibliographystyle{IEEEtran}
\balance
\bibliography{IEEEabrv,references}

\end{document}